\documentclass{PoS}

\title{Some nucleon isovector observables from 2+1-flavor domain-wall QCD at the physical pion mass}

\ShortTitle{Nucleon structure in 2+1-flavor DWF QCD}

\author{\speaker{Shigemi Ohta} 
for LHP and RBC and UKQCD Collaborations\\
        Theory Center, KEK, Tsukuba, Ibaraki 305-0801, Japan\\
        Department of Particle and Nuclear Physics, SOKENDAI, Hayama, Kanagawa, 2400193, Japan\\
        RIKEN BNL Research Center, BNL, Upton, NY 11973, USA\\
        E-mail: \email{shigemi.ohta@kek.jp}}

\abstract{
The current status of the LHP and RBC joint calculations of the nucleon isovector form factors and low moments of structure functions with a 2+1-flavor dynamical domain-wall fermion (DWF) lattice-QCD ensemble at the physical pion mass generated by RBC and UKQCD Collaborations with a momentum cutoff of 1.730(4) GeV and lattice spatial extent of 5.476(12) fm is reported.
About ten percent of the statistics reported in Lattice 2014 were found with an incorrect boundary condition in time but correcting for it resulted in less than one-percent difference.

\vspace{-169mm}\parbox{\textwidth}{\flushright\large\rm \hfill KEK-TH-1871, RBRC-1150}\vspace{162mm}
}

\FullConference{The 33rd International Symposium on Lattice Field Theory\\
		14 -18 July 2015\\
		Kobe International Conference Center, Kobe, Japan*}

\begin{document}

\section{Introduction}

As was reported in Lattice 2014 \cite{Syritsyn:2014xwa}, the Lattice Hadron Physics (LHP) and RIKEN-BNL Columbia (RBC) Collaborations began a joint research on nucleon structure using the physical-mass dynamical 2+1-flavor domain-wall fermion (DWF) lattice-QCD ensembles jointly generated by the RBC and UKQCD Collaborations \cite{Blum:2014tka}.
The initial calculations are done on the coarser of the two ensembles at an inverse lattice spacing, \(a^{-1}\) of about 1.730(4) GeV/c, linear extent of the lattice in spatial directions of 48 or about 5.476(12) fm, and a pion mass of about 139.2(4) MeV.
We used 20 configurations that evenly span the entire ensemble in equilibrium of about 1,500 trajectories.
The nucleon source was optimized in much the same way as in the previous RBC nucleon calculations \cite{Ohta:2014rfa}.
In addition we use four different source-sink separations of 8, 9, 10, and 12 units that span the range of about 0.9 to 1.4 fm so we should characterize and remove whatever small leftover excited-state contamination:
When we decide the lattice action, the mass spectra of hadrons, \(E_0, E_1, ...\), are decided.
As we decide on source and sink smearings, amplitudes of each ground and excited states are defined, \(e^{-E_0 t} |0\rangle + A_1 e^{-E_1t}|1\rangle + ...\)
Though we try to optimize so excited-state amplitudes such as \(A_1\) are small, some dependence on source-sink separation, \(t_{\rm sep}=t_{\rm sink}-t_{\rm source}\), remains:
\(
\langle 0| O |0\rangle + A_1 e^{-(E_1-E_0)t_{\rm sep}} \langle 1|O|0\rangle + ...
\)
unless the observable is a conserved charge, \(Q\), because \(\langle 1|Q|0\rangle = 0\).
Indeed we are yet to detect such dependence in both vector and axial charges, but have detected it in quark momentum and helicity fractions.
By simultaneously collecting data on the source-sink separation dependence of many observables we plan to determine the excitation energy \(E_1-E_0\), amplitude \(A_1\), and off-diagonal matrix elements \(\langle 1|O|0\rangle\) as well as the ground-state values \(\langle 0| O |0\rangle\) that we seek.

To efficiently increase the statistics we use the AMA covariant approximation average method \cite{Shintani:2014vja} in combination with a low-mode deflation.
Unfortunately the deflation could not be done to the desired level because of memory-size limitations in available computing resources.
To compensate we would have to increase the number of configurations by decreasing the intervals between them.
So we planned accordingly, but could not gain any more statistics during the past year because the eligible computers were all over-subscribed.
Thus this report would have been pointless except one: we found some errors in our handling of the timelike boundary condition in about ten percent of our AMA samples.
Fortunately correcting for this error did not result in any correction that is larger than one percent in any of the observables we reported last year.

\section{Nucleon structure in domain-wall QCD}

The RBC and joint RBC+UKQCD collaborations have been investigating nucleon structure using the domain-wall fermions (DWF) quarks \cite{Sasaki:2003jh,Orginos:2005uy,Lin:2008uz,Yamazaki:2008py,Yamazaki:2009zq,Aoki:2010xg,Lin:2014saa,Ohta:2014rfa} on a sequence of 0-, 2- and 2+1-flavor dynamical DWF ensembles at various mass values \cite{Blum:2000kn,Aoki:2004ht,Allton:2008pn,Aoki:2010dy,Arthur:2012opa,Blum:2014tka}.
The LHP collaboration also calculated some nucleon structure \cite{Syritsyn:2009mx} using a RBC+UKQCD 2+1-flavor dyamical DWF ensemble \cite{Allton:2008pn}.
As is well known, the DWF scheme allows to maintain continuum-like flavor and chiral symmetries on the lattice, and helps to simplify non-perturbative renormalizations.
These RBC and RBC+UKQCD nucleon calculations exposed persistent underestimates of the isovector axial charge, by about ten percent, as compared with the experiment, \(g_A/g_V = 1.2723(23)\) \cite{Agashe:2014kda}.
This has been confirmed by many other calculations with different fermion schemes \cite{Constantinou:2014tga}.
As the isovector axial charge not only determines the neutron life but also the pion-nucleon coupling through Goldberger-Treiman relation, this underestimation may mean the entire nuclear physics is not well described by conventional lattice QCD.
The problem is associated with unusually long-range autocorrelation of this obsevable seen only with the smallest values of \(m_\pi L\) finite-size scaling parameter \cite{Ohta:2013qda,Ohta:2014rfa}.
This suggests satisfactory calculation of nucleon structure may require significantly larger lattice spatial volume than conventional ones with \(m_\pi L\sim 4\) for light-meson observables.
Another persistent difficulty is the overestimate of isovector quark momentum and helicity fractions  \cite{Constantinou:2014tga}.
Here the most recent RBC+UKQCD work captured an encouraging trend toward the experiment in their lighter-mass DWF calculations \cite{Aoki:2010xg,Ohta:2014rfa}.
Both these difficulties point to a need for 2+1-flavor dynamical DWF calculations of nucleon structure at lighter mass values  and larger spatial volumes.
Fortunately now such lighter-mass ensembles exist, indeed at physical mass \cite{Blum:2014tka}, albeit with small spatial volumes.
Thus another joint collaboration of LHP and RBC was formed to explore nucleon structure on these physical-mass 2+1-flavor DWF ensembles \cite{Syritsyn:2014xwa}.

\section{Problems 2014-2015}

During the past year this new LHP+RBC joint collaboration encountered two problems that prevented them from accumulating more statistics: one is the mishandling of the timelike boundary condition in part of their AMA samples and the other is the general oversubscription of computers capable of running their calculations.
The former was discovered soon after Lattice 2014.
We use periodic+antiperiodic boundary condition for the fermions in the time direction of the lattice that is twice longer than the space directions.
This only matters when the boundary falls upon between the nucleon source and sink.
In our AMA calculations, the source position of the one precise sample on each configuration moves with a constant displacement vector of (17, 17, 17, 17) from one configuration to the next.
The 32 imprecise samples per configuration are displaced from this by half the lattice spatial extent in all directions including time.
Thus slightly less than one in eight configurations risk the error in handling the timelike boundary condition, also dependent on the nucleon source-sink separations.
About ten percent of the samples reported last year were contaminated this way.
Even with this error present, the difference from calculations without the error should be minor because it is very much like using only-slightly non-unitary valence quarks.
The corrected results prove that the error was minor indeed.
Correcting for this was easy and did not take too long either, but nonetheless lost some crucial computing resources that could have been used for accumulating more statistics.
\section{Corrected Results}

\begin{figure}[h]
\includegraphics[width=0.48\textwidth]{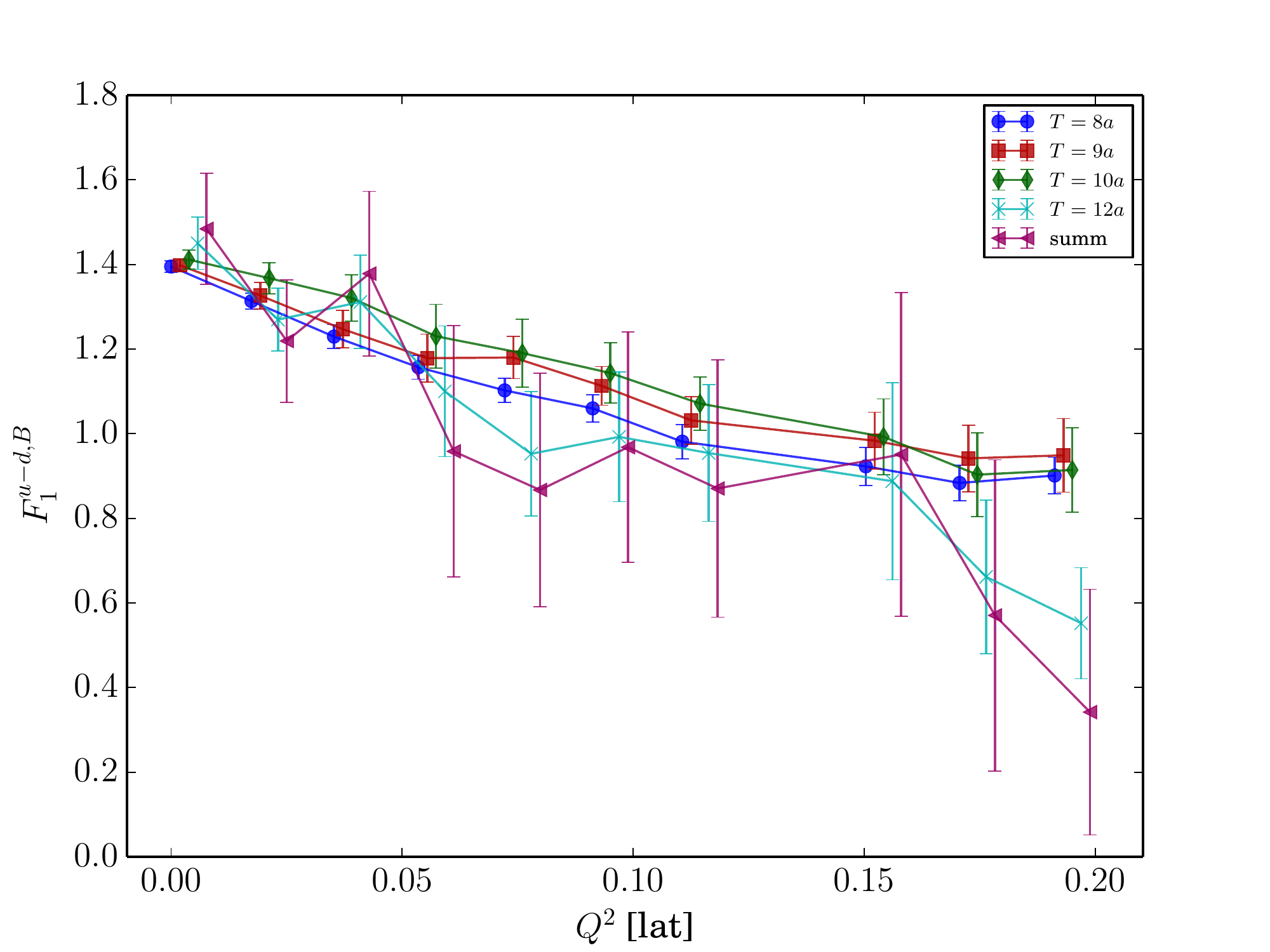}
\includegraphics[width=0.48\textwidth]{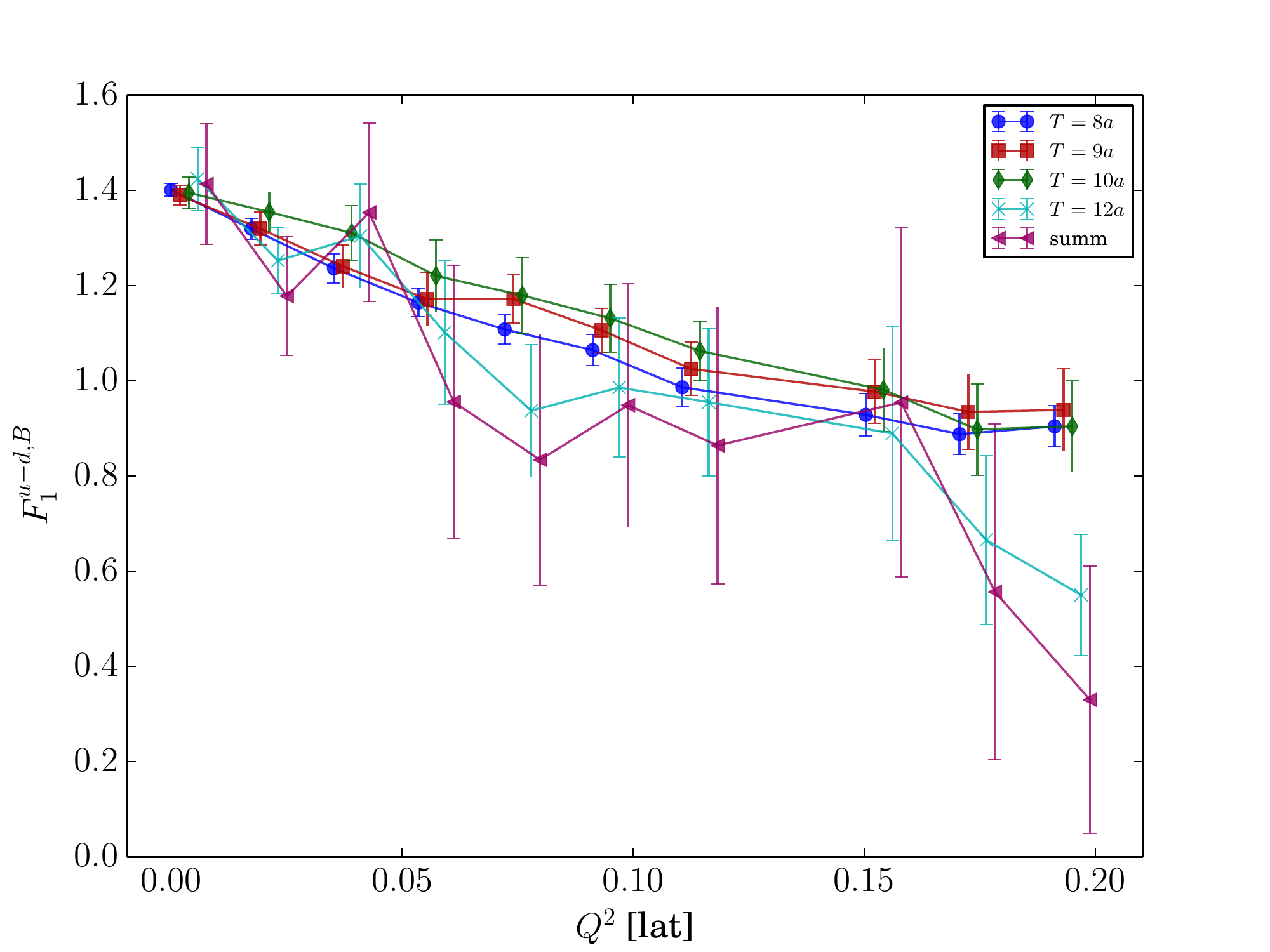}
\caption{\label{fig:F1}
Nucleon isovector Dirac form factor, \(F_1\), before (left) and after (right) correcting for the timelike boundary condition.
Correction results in only less than 1-\% difference.
}
\end{figure}

\begin{figure}[h]
\includegraphics[width=0.48\textwidth]{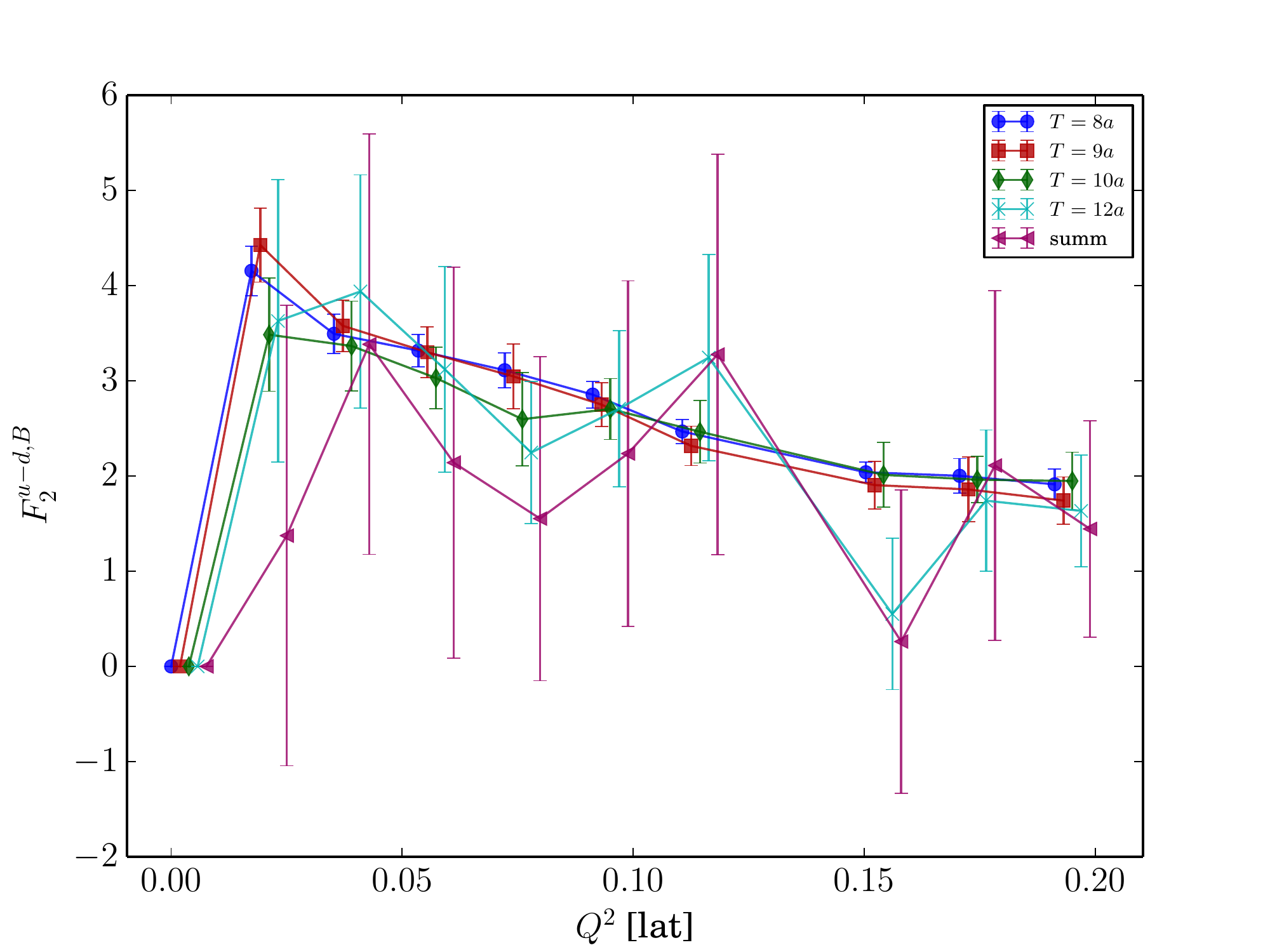}
\includegraphics[width=0.48\textwidth]{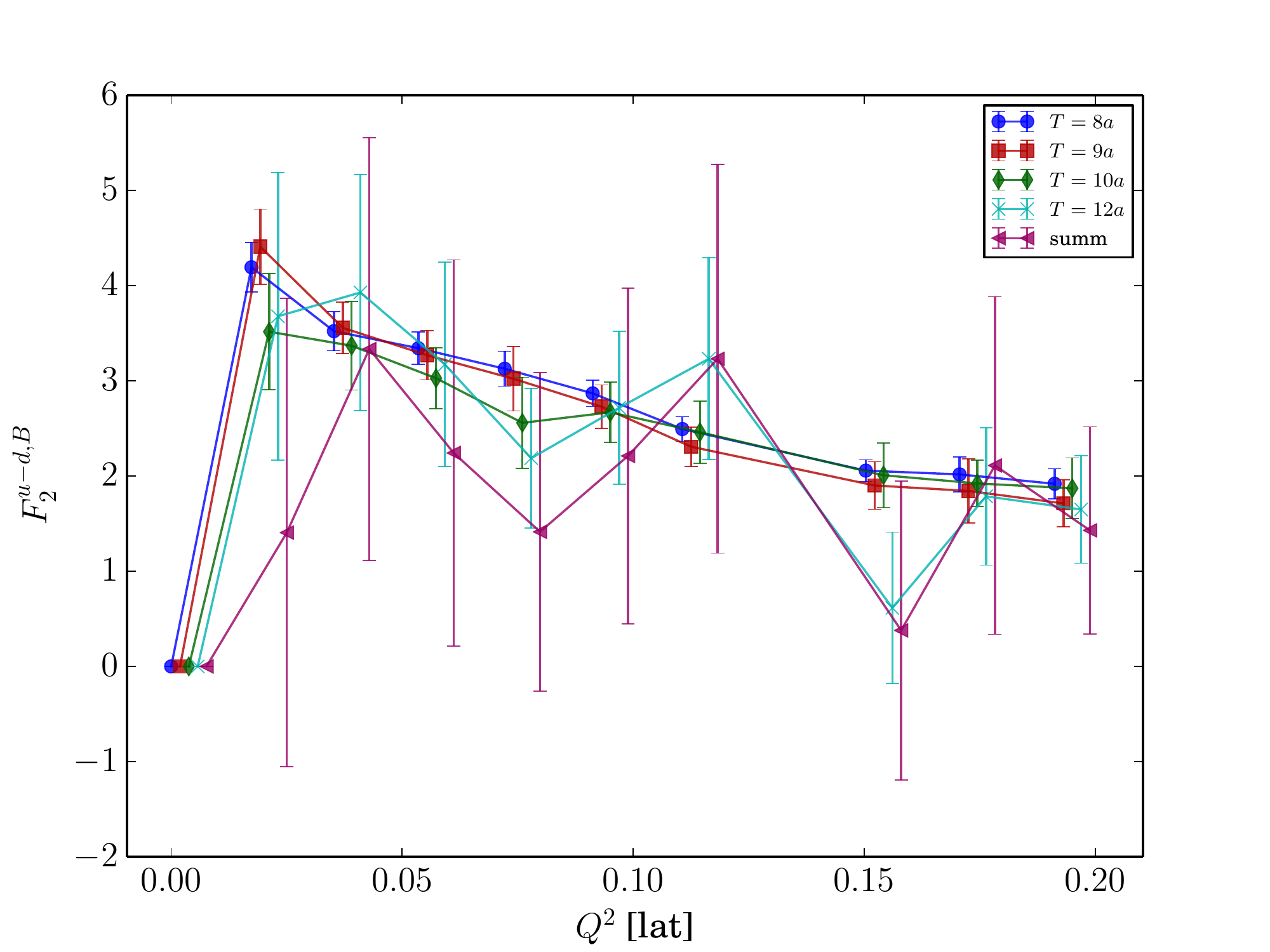}
\caption{\label{fig:F2}
Nucleon isovector Pauli form factor, \(F_2\), before (left) and after (right) correcting for the timelike boundary condition.
Correction results in only less than 1-\% difference.
}
\end{figure}

\begin{figure}[h]
\includegraphics[width=0.48\textwidth]{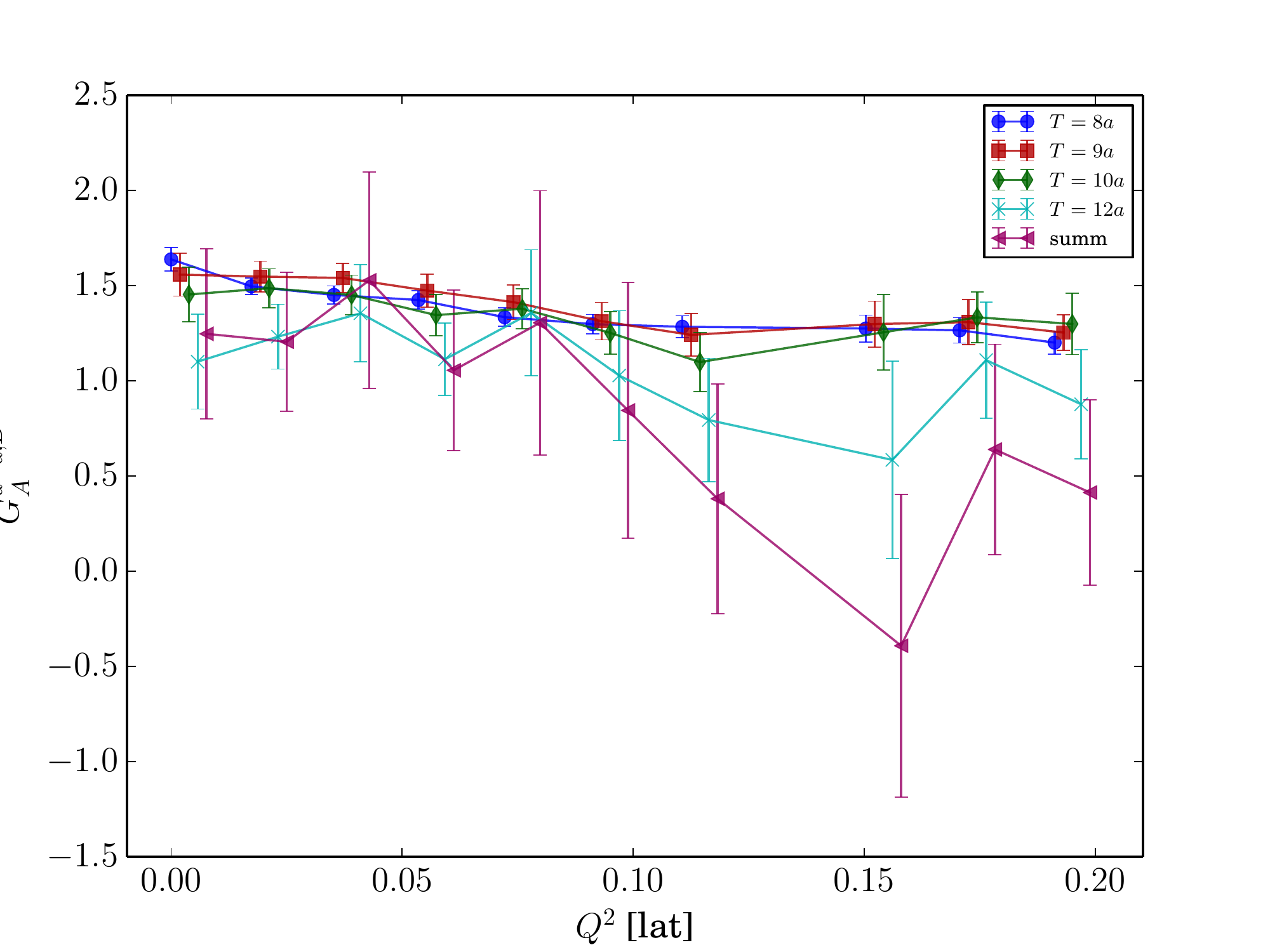}
\includegraphics[width=0.48\textwidth]{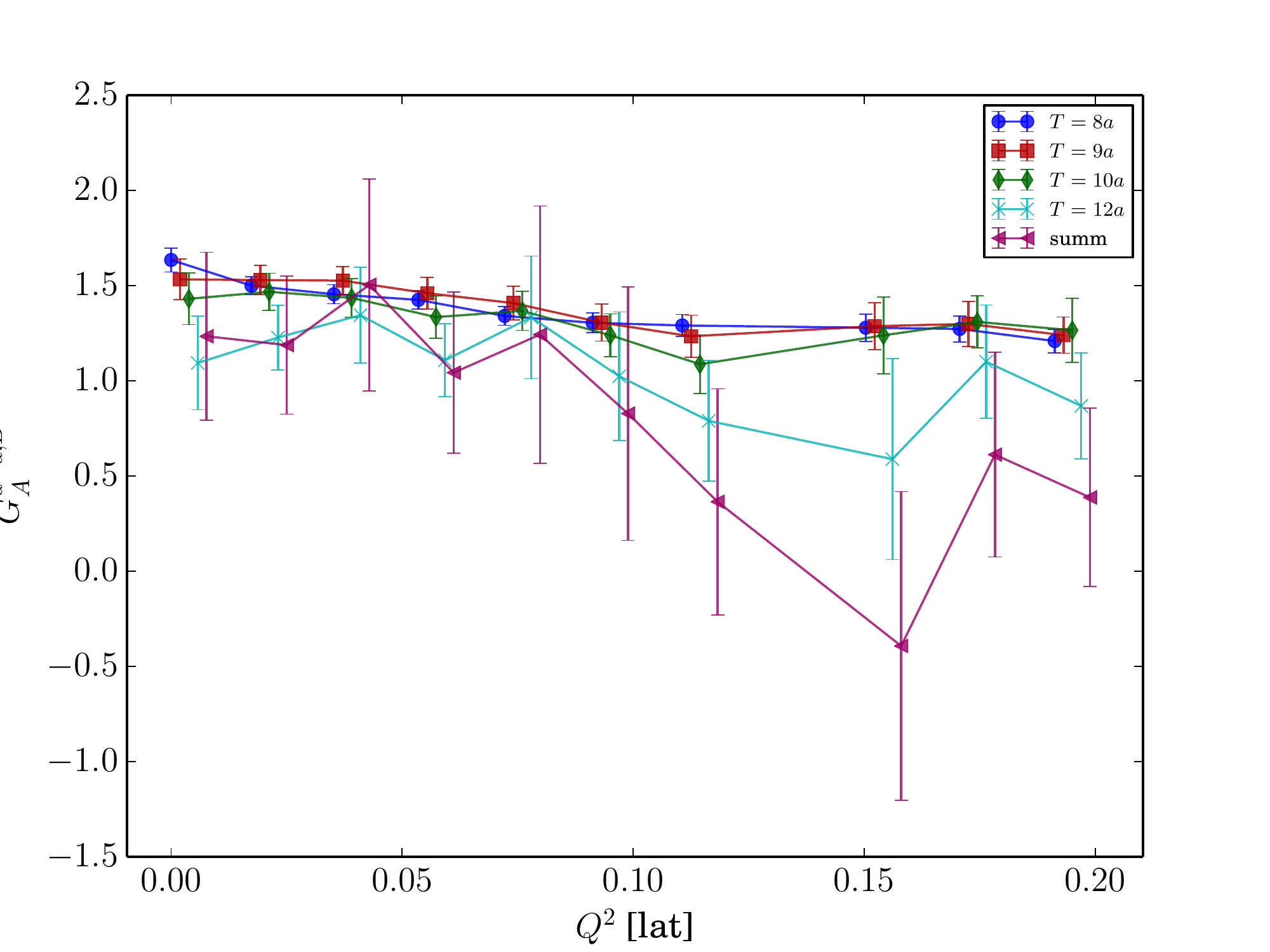}
\caption{\label{fig:GA}
Nucleon isovector axialvector factor, \(G_A\), before (left) and after (right) correcting for the timelike boundary condition.
Correction results in only less than 1-\% difference.
}
\end{figure}

\begin{figure}[h]
\includegraphics[width=0.48\textwidth]{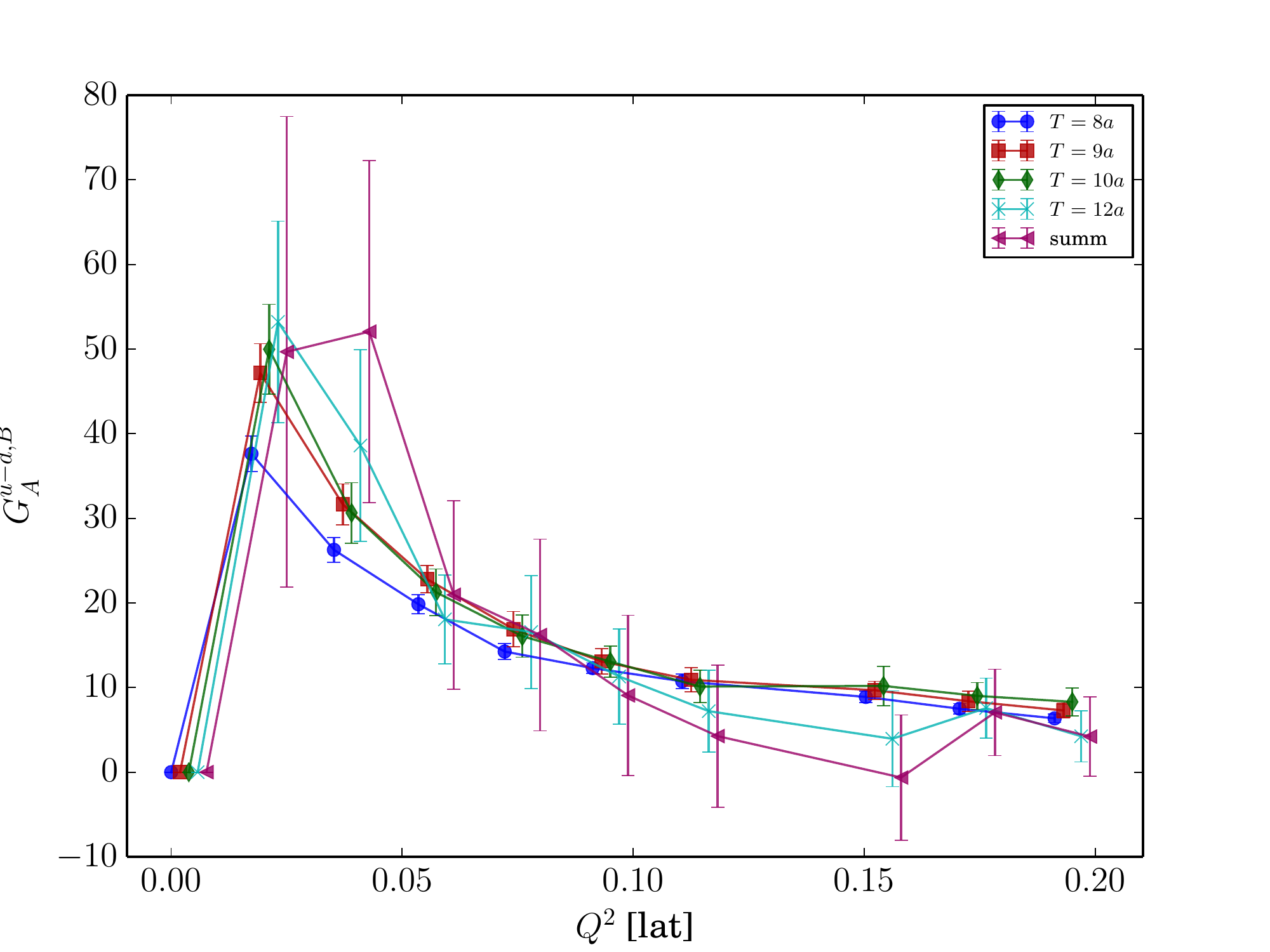}
\includegraphics[width=0.48\textwidth]{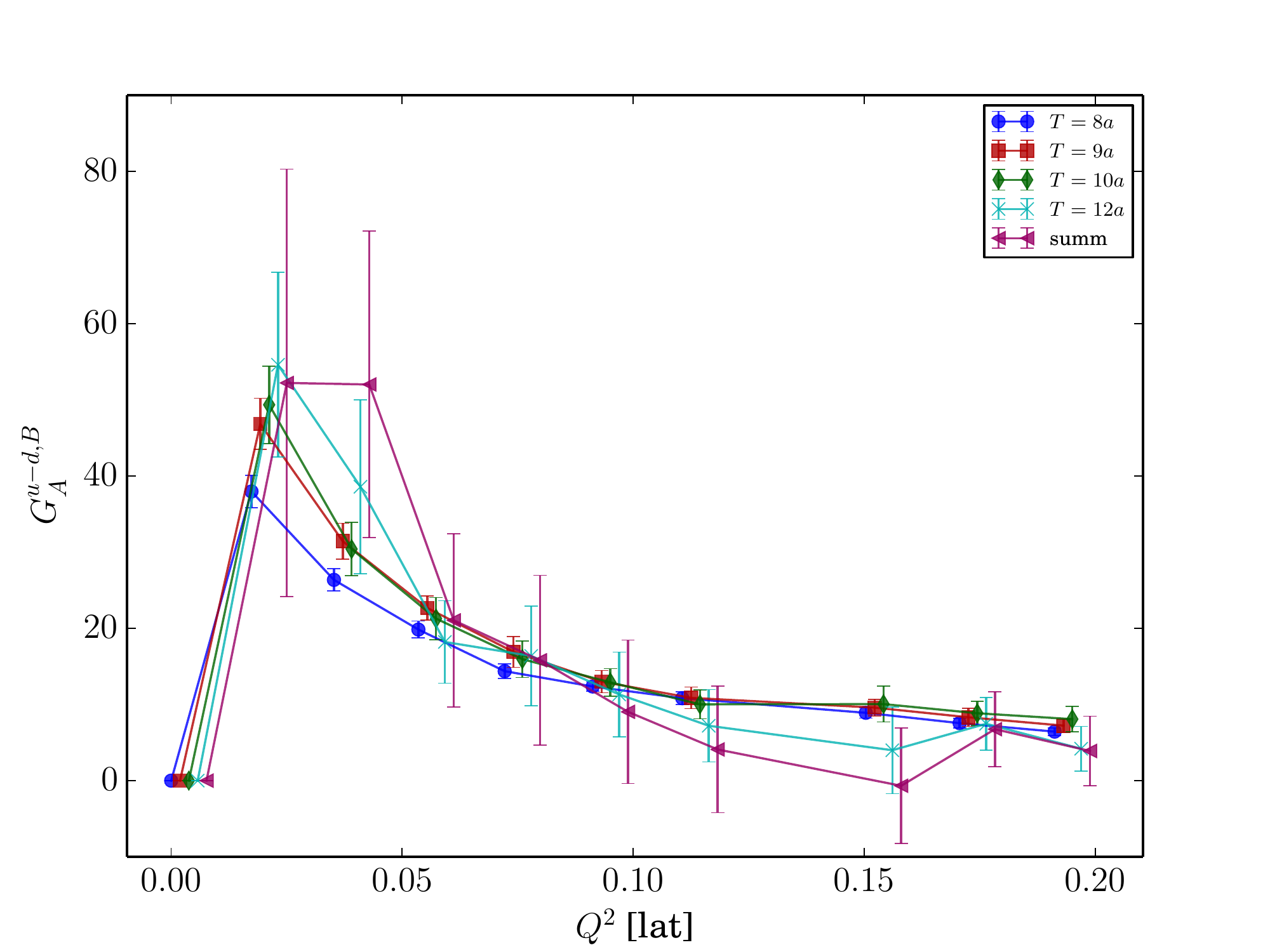}
\caption{\label{fig:GP}
Nucleon isovector pseudoscalar form factor, \(G_P\), before (left) and after (right) correcting for the timelike boundary condition.
Correction results in only less than 1-\% difference.
}
\end{figure}

\begin{figure}[h]
\includegraphics[width=0.48\textwidth]{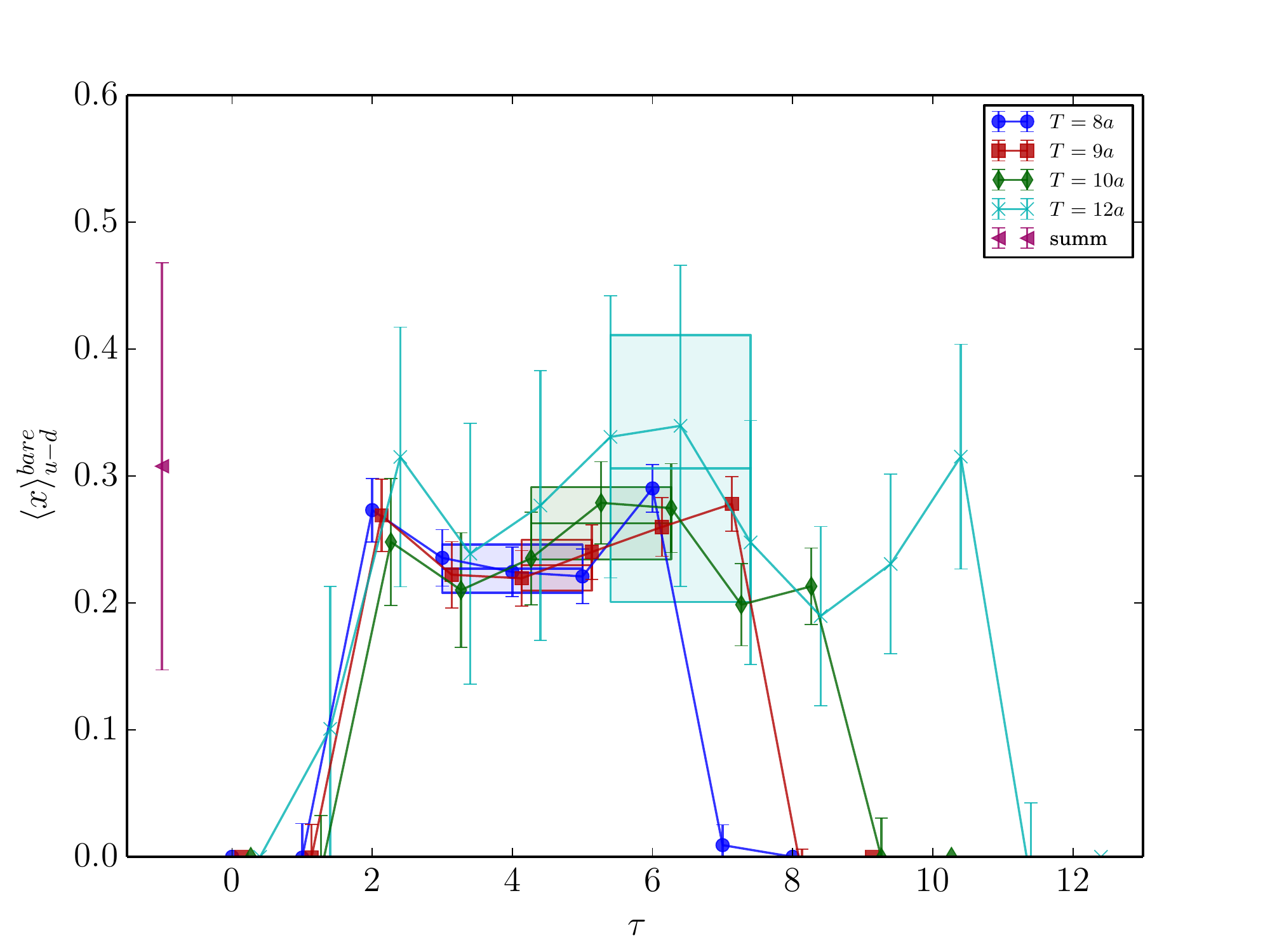}
\includegraphics[width=0.48\textwidth]{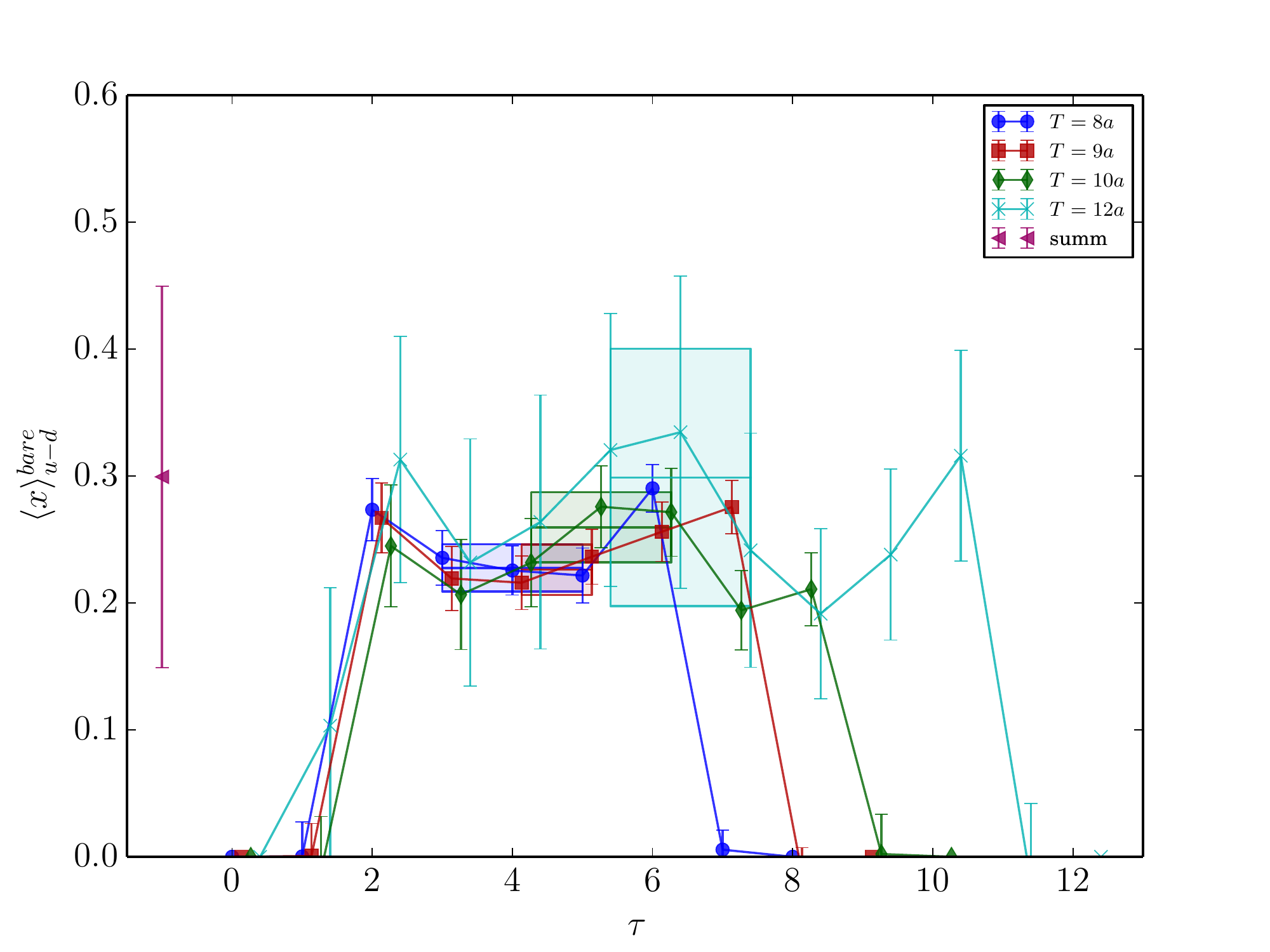}
\caption{\label{fig:xq}
Nucleon isovector quark momentum fraction, \(\langle x \rangle_{u-d}\), before (left) and after (right) correcting for the timelike boundary condition.
Correction results in only less than 1-\% difference.
}
\end{figure}

Here are some representative isovector observables before and after correcting for the mishandling of timelike boundary condition.
First, the isovector vector charge, \(g_V\), in Fig.\ \ref{fig:gV}:
\begin{figure}[t]
\includegraphics[width=0.48\textwidth]{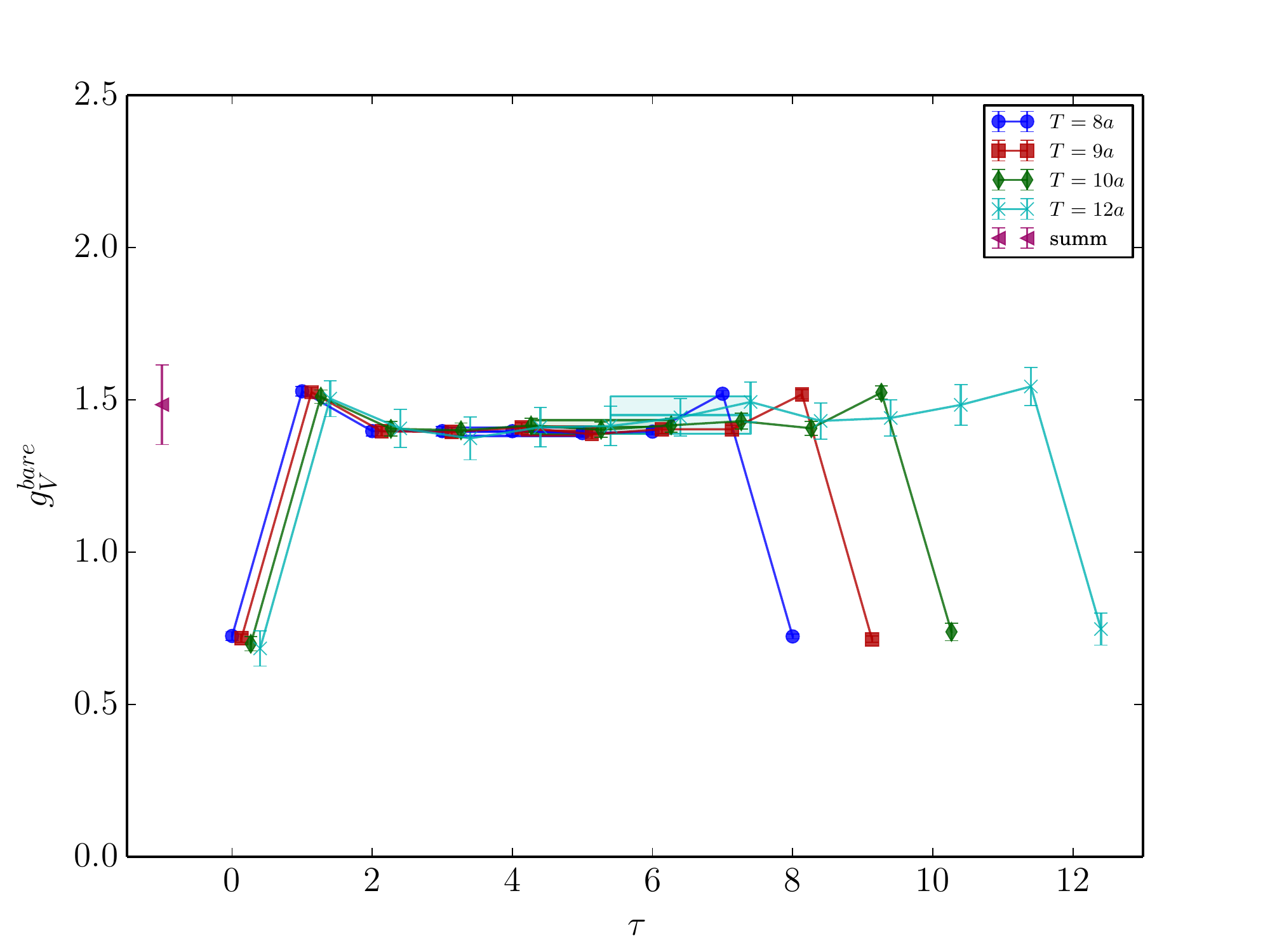}
\includegraphics[width=0.48\textwidth]{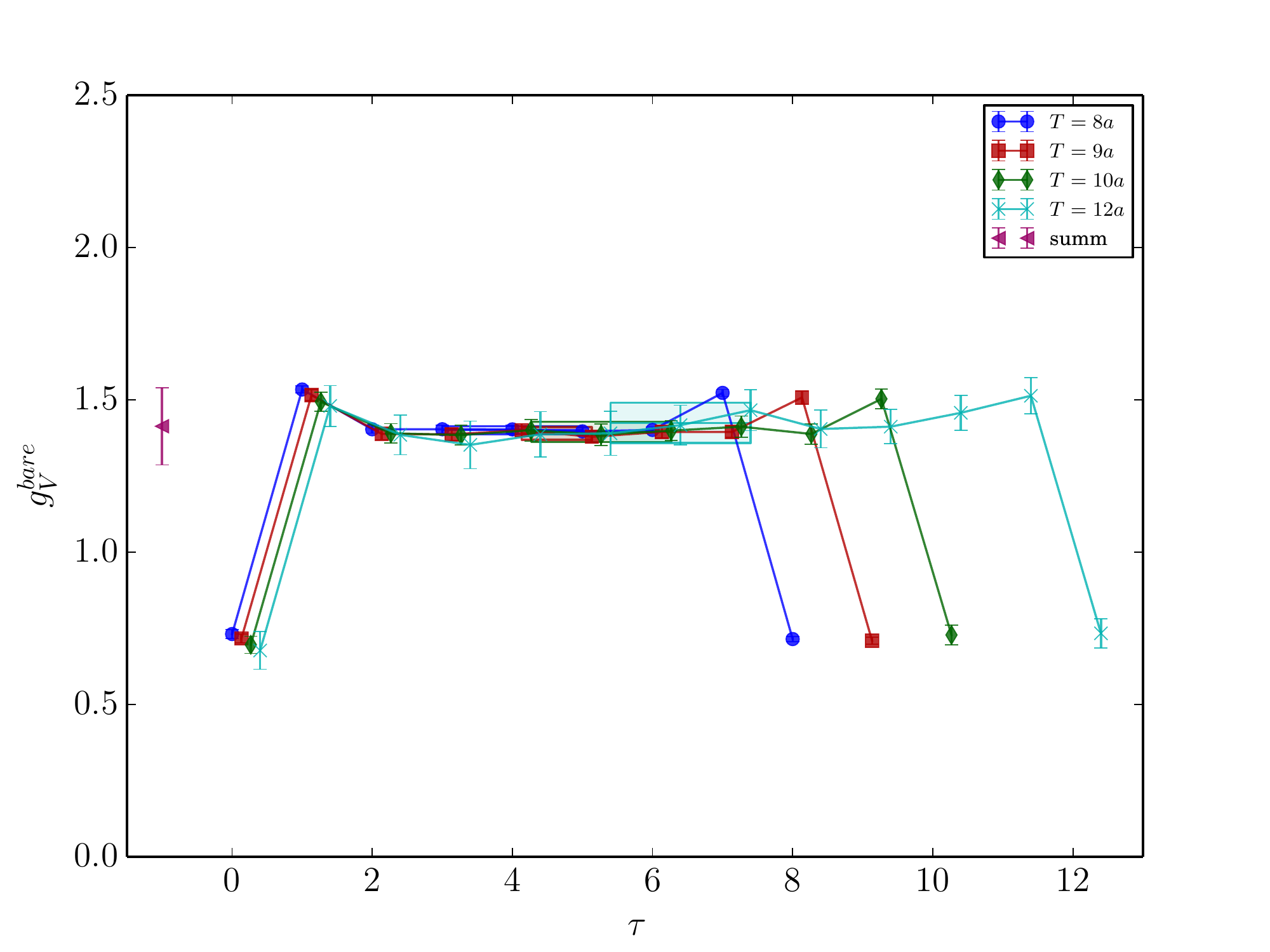}
\caption{\label{fig:gV}
Nucleon isovector vector charge, \(g_V\), before (left panel) and after (right) correcting for the error in timelike boundary-condition.
Correction results in only less than 1-\% difference.
}
\end{figure}
The old results with the error is presented in the left panel, and the corrected result is in the right.
The correction results in less than one-percent difference.
Second, the isovector axial charge, \(g_A\), in Fig.\ \ref{fig:gA}:
\begin{figure}[h]
\includegraphics[width=0.48\textwidth]{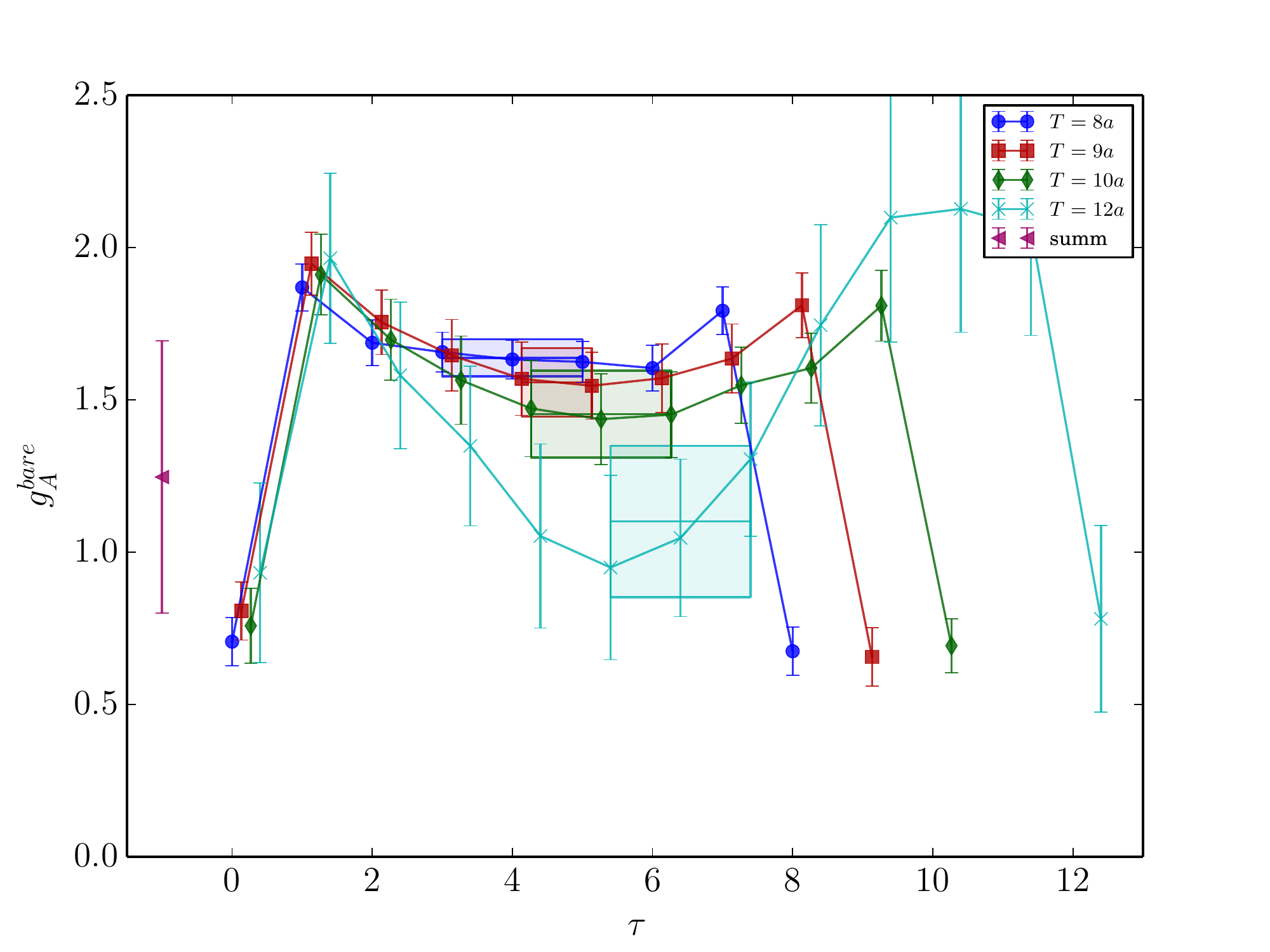}
\includegraphics[width=0.48\textwidth]{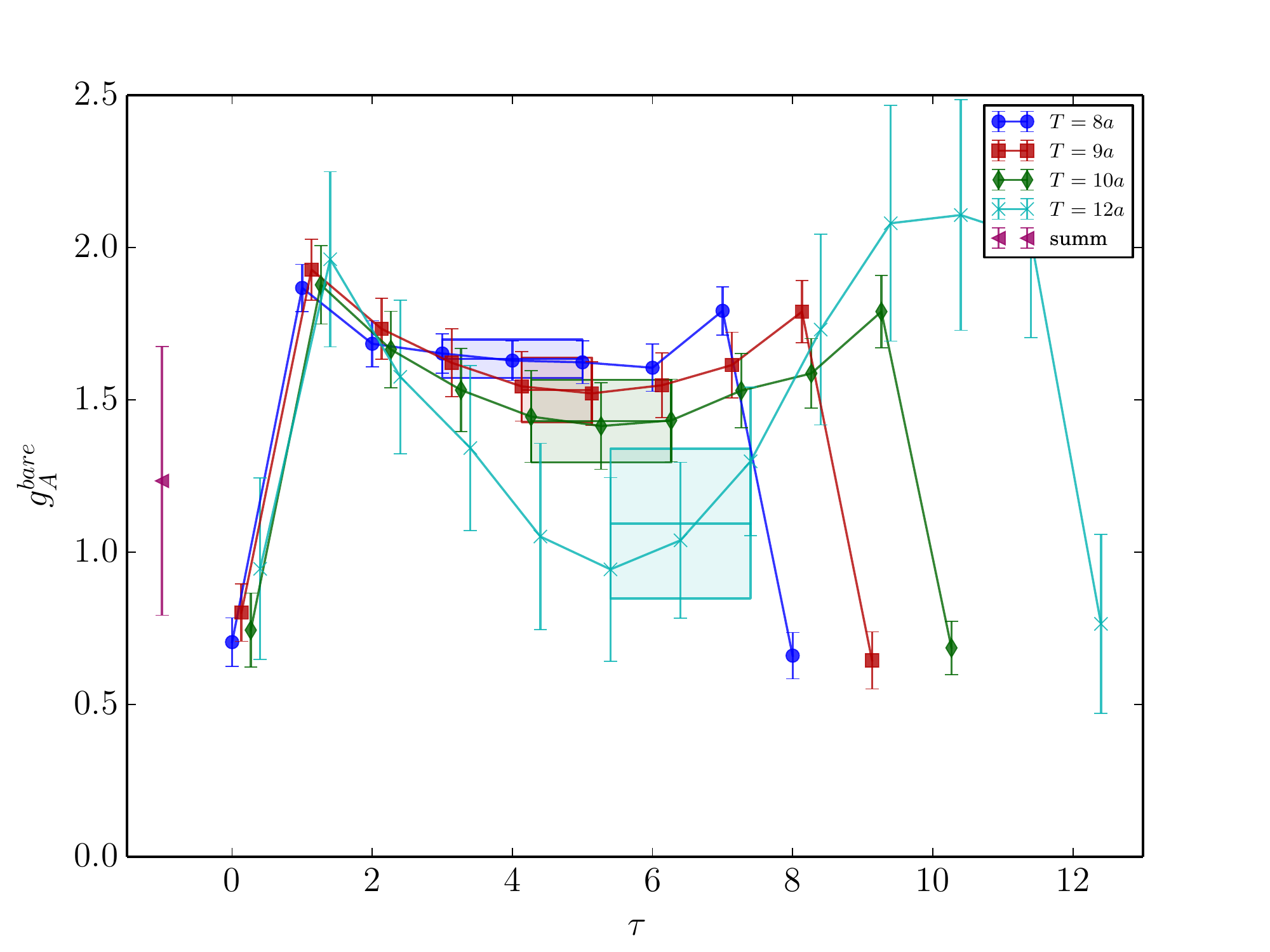}
\caption{\label{fig:gA}
Nucleon isovector axial charge, \(g_A\), before (left) and after (right) correcting for the timelike boundary condition.
Correction results in only less than 1-\% difference.
}
\end{figure}
again the correction results in less than one-percent difference.
Third, the ratio, \(g_A/g_V\), of isovector axial and vector charges, in Fig.\ \ref{fig:gAgV}:
\begin{figure}[h]
\includegraphics[width=0.48\textwidth]{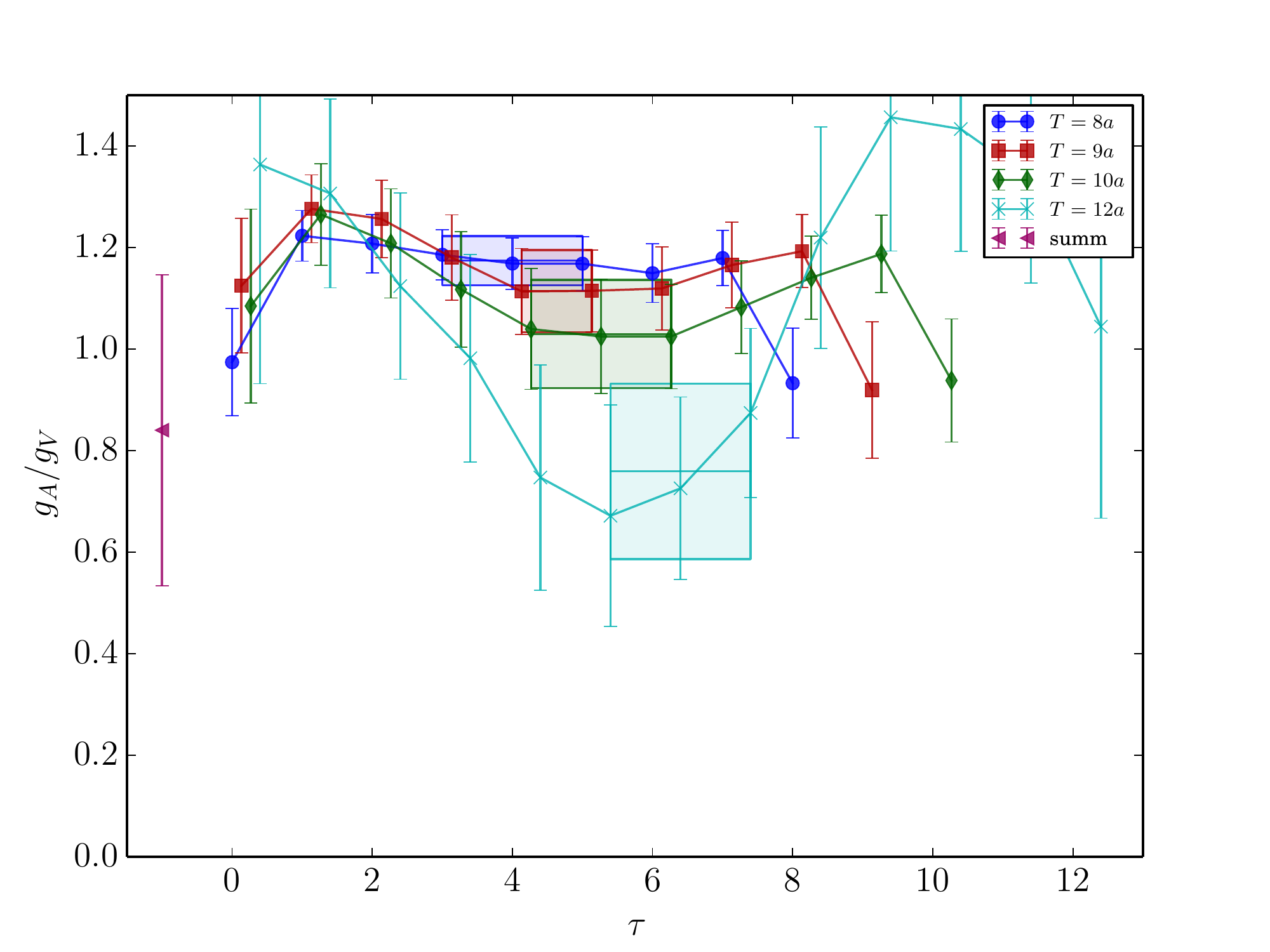}
\includegraphics[width=0.48\textwidth]{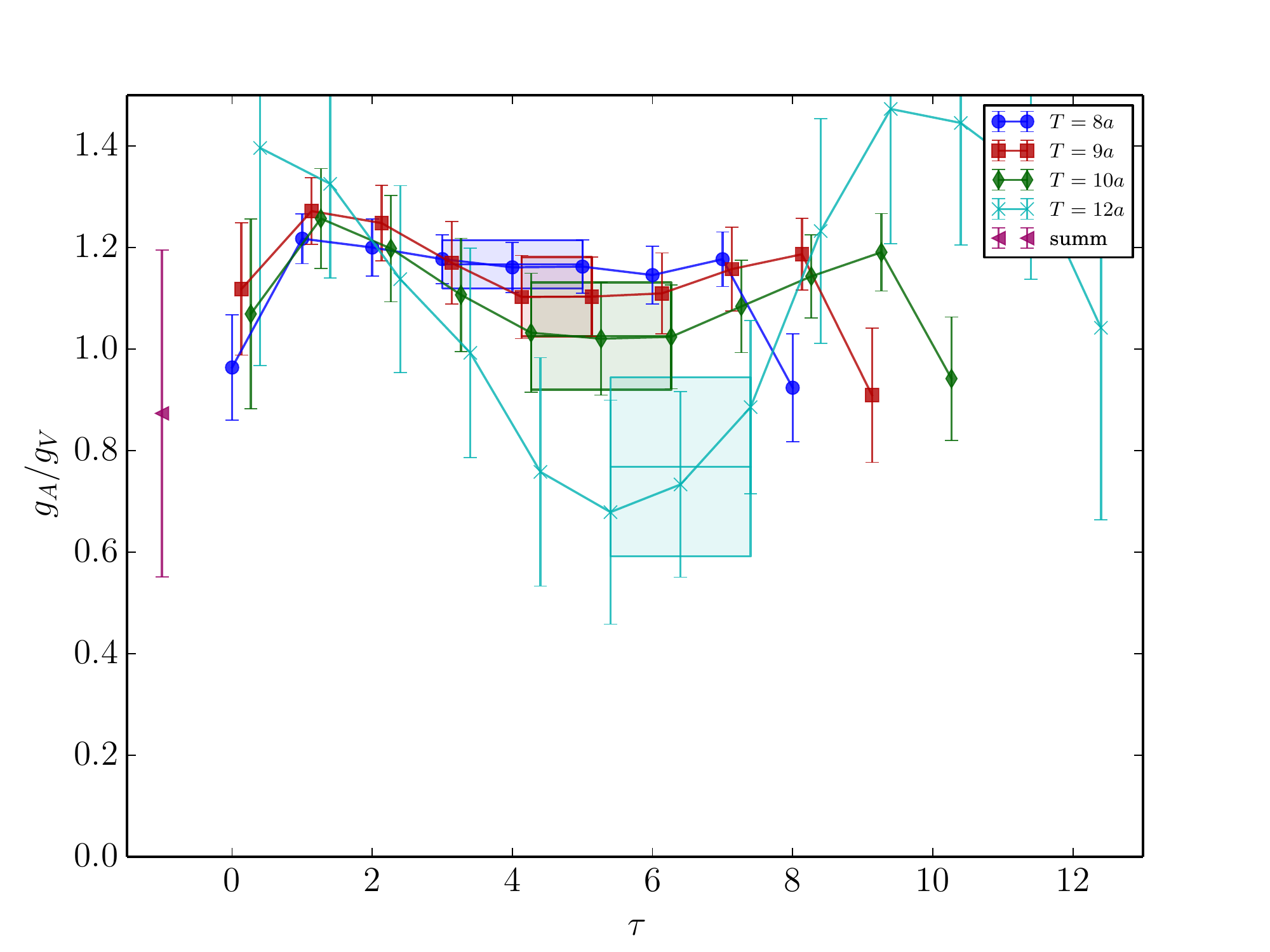}
\caption{\label{fig:gAgV}
Ratio of nucleon isovector axial and vector charge, \(g_A/g_V\), before (left) and after (right) correcting for the timelike boundary condition.
Correction results in only less than 1-\% difference.
}
\end{figure}
yet again the correction results in less than one-percent difference.
In all the cases, the Dirac form factor, \(F_1\), in Fig.\ \ref{fig:F1},
Pauli form factor, \(F_2\), in Fig.\ \ref{fig:F2},
Axial form factor, \(G_A\), in Fig.\ \ref{fig:GA},
Pseudoscalar form factor, \(G_P\), in Fig.\ \ref{fig:GP},
and Quark momentum fraction, \(\langle x \rangle_{u-d}\), in Fig.\ \ref{fig:xq},
the correction results in less than one-percent differences.

As can be seen from these corrected results, no isovector nucleon observable was found affected for more than a percent by the mishandling of the timelike boundary condition that affected about ten percent of the AMA samples.
This is easy to understand as the difference from calculations without the error should be minor because it is very much like using only-slightly non-unitary valence quarks.
The corrected results prove that the error was minor indeed.

However, as can also be seen, we need a lot more statistics to draw any physically useful conclusion from the corrected results.

\section{Conclusions}

About ten percent of our AMA samples reported in Lattice 2014 were affected by our mishandling of the timelike boundary condition.
Correcting for this was easy and did not take too long either.
Even with this error present, the difference from calculations without the error should be minor because it is very much like using only-slightly non-unitary valence quarks.
The corrected results prove that the error was minor indeed.
Nonetheless we lost some crucial computing resources that could have been used for accumulating more statistics.
We are eager to recover this loss by taking advantage of newer and better computers that are now available.
We will be able to adjust our low-mode deflation details for the better and faster calculations.

It is a great pleasure to thank close and enjoyable collaboration with S.~Syritsyn, T.~Blum, T.~Izubuchi, C.~Jung, M.~Lin and E.~Shintani, and with past and present members of the LHP and RBC collaborations who contributed nucleon matrix elements algorithms and their installations, and past and present members of the RBC and UKQCD collaborations who contributed the DWF ensembles, in particular the 2+1-flavor 1.73-GeV physical mass ensemble.
This ensemble was generated using the IBM Blue Gene/Q (BG/Q) “Mira” machines at the Argonne Leadership Class Facility (ALCF) provided under the Incite Program of the US DOE, on the STFC funded “DiRAC” BG/Q system in the Advanced Computing Facility at the University of Edinburgh, and on the BG/Q machines at Brookhaven National Laboratory (BNL).
Computations of the nucleon matrix elements were carried out on facilities of the USQCD Collaboration, which are funded by the Office of Science of the U.S. Department of Energy.

\end{document}